# Intra-wire coupling in segmented Ni/Cu nanowires deposited by electrodeposition

Philip Sergelius[1], Ji Hyun Lee[2], Olivier Fruchart[5,6], Mohamed Shaker Salem[1,8], Sebastian Allende[3], Roberto Alejandro Escobar[3], Johannes Gooth[1,4], Robert Zierold[1], Jean-Christophe Toussaint[5], Sebastian Schneider[10], Darius Pohl[10], Bernd Rellinghaus[10], Sylvain Martin[5], Javier Garcia[1, 10], Heiko Reith[1,10], Anne Spende[7,9], Maria-Eugenia Toimil-Molares[7], Dora Altbir[3], Russel Cowburn[2], Detlef Görlitz[1], Kornelius Nielsch[1, 10]

*[1] Institute of Nanostructure and Solid-State Physics, Universität Hamburg, 20355 Hamburg, Germany*
*[2] Department of Physics, University of Cambridge, Cambridge CB3 0HE, United Kingdom*
*[3] Departamento de Física, CEDENNA, Universidad de Santiago de Chile, USACH, 9170124 Santiago, Chile*
*[4] IBM Research GmbH, 8803 Rueschlikon, Switzerland*
*[5] Université Grenoble Alpes, CNRS, Institut Néel, 38042 Grenoble, France*
*[6] Université Grenoble Alpes, CNRS, CEA, SPINTEC, 38054 Grenoble, France*
*[7]GSI Helmholtzzentrum für Schwerionenforschung GmbH, 64291 Darmstadt, Germany*
*[8] Physics Department, Faculty of Science, Cairo University, 12613 Giza, Egypt*
*[9] Material- und Geowissenschaften, Technische Universität Darmstadt, 64287 Darmstadt, Germany*
*[10] Leibnitz Institute for Solid State and Materials Research (IFW), 01069 Dresden, Germany*

**ABSTRACT**

Segmented magnetic nanowires are a promising route for the development of three dimensional data storage techniques. Such devices require a control of the coercive field and the coupling mechanisms between individual magnetic elements. In our study, we investigate electrodeposited nanomagnets within host templates using vibrating sample magnetometry and observe a strong dependence between nanowire length and coercive field (25 nm to 5 µm) and diameter (25 nm to 45 nm). A transition from a magnetization reversal through coherent rotation to domain wall propagation is observed at an aspect ratio of approximately 2. Our results are further reinforced *via* micromagnetic simulations and angle dependent hysteresis loops. The found behavior is exploited to create nanowires consisting of a fixed and a free segment in a spin-valve like structure. The wires are released from the membrane and electrically contacted, displaying a giant magnetoresistance effect that is attributed to individual switching of the coupled nanomagnets. We develop a simple analytical model to describe the observed switching phenomena and to predict stable and unstable regimes in coupled nanomagnets of certain geometries.

1. INTRODUCTION

In the investigation of nanomagnet assemblies, the focus has mainly been on lithographically-patterned, planar structures, which can be used as logic devices capable of executing Boolean logic operations [1-8]. Often applications are sought within high density recording, bit patterned media [9-12] or MRAM devices [1,13], however since the bit size of modern hard disk drives (approx. 20x20 nm²) is already lower than what is achievable with most lithography techniques [14], potential use is limited. Therefore, the use of the third dimension is an appealing approach. Studies on alternating, vertical stacks of magnetic and non-magnetic materials have already been published, in which a three dimensional shift register is investigated, that is capable of storing and shifting several bits of information as solitons within a single column [15-19]. These shift registers are coupled *via* RKKY interaction and its oscillatory nature allows for an additional degree of freedom in the design of the device, but as a drawback the samples are very demanding with regard to sample quality and purity. A soliton, which is the bearer of information, is a magnetic frustration within a 1D chain of magnetic moments. It is thinkable to induce such a soliton using dipolar coupling instead of RKKY, simply by magnetizing two nanomagnets in the same direction and placing them next to each other (transversal soliton [20]). Another possibility would be to magnetize them in opposing direction and place them in a head-to-head or tail-to-tail configuration (longitudinal soliton, Figure 1). The coupling strength is then defined by the material properties and the distance of both nanomagnets. Such kinds of nanowires can be synthesized within porous anodic aluminum oxide (AAO) templates [21-23] using electrodeposition. In general, electrodeposition within porous materials is a favorable technique for synthesis of three dimensional structures, because no other technique is capable of obtaining large and dense arrays, where each individual wire has an aspect ratio larger than 1000 and still can be modulated along its axis of growth. We chose voltage-dependent electrodeposition from a single electrolyte containing a high concentration of Ni ions for the magnetic segments and a low concentration of Cu ions for the non-magnetic segments [24-30]. Other possible modulation techniques are variations of the diameter [23,31] or crystal structure [32,33]. It has also been shown that AAO templates can have perfect order [34], which is crucial for applications, and with usage of entirely different templates even the cross-sectional pore geometry can be tuned [35,36]. We chose nickel as a magnetic material, because it is chemically robust against oxidation and has the lowest magnetic moment of the three 3d-ferromagnets. Lower magnetic moments are equivalent with lower stray fields, which allows us to narrow the gap between two magnetic segments and to reduce the dipolar interaction between neighboring wires [37,38].

For a successful application of coupled nanomagnets within nanowire storage arrays, the coercive field needs to be tunable, because specific segments need to be able to change their direction of magnetization, while others remain magnetized as before. In this publication, we will first discuss the synthesis of pure Ni nanomagnets with different diameters and lengths, followed by a discussion of the synthesis of coupled nanomagnets. The individual properties of the nanomagnets will be investigated first and based on these results, an assessment of the coupling mechanisms and strength will follow. The individual switching of single nanomagnets will be evidenced using electric transport measurements displaying a giant magnetoresistance effect. In the end an analytical model will be presented describing the observed phenomena.

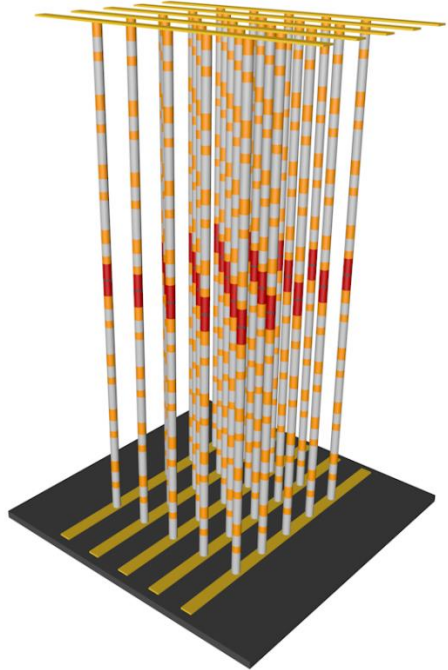

**Figure 1** Longitudinal soliton shift register based on dipolar coupling consisting of magnetic segments (gray), copper segments (orange), a read element (red) and read-lines (gold top/bottom). In the read process, magnetic information is propagated up- or downwards through the read element, whereas a writing process can be conducted from the top or bottom.

2. **EXPERIMENTAL SECTION**

*2.1 Synthesis of individual nanomagnets.* Commercial, highly ordered AAO templates (SmartMembranes GmbH) with an interpore distance of 105 nm and pore diameters of 35 nm and 60 nm were used. For a diameter dependent study, the pore diameters of the templates were reduced to 25 nm and 45 nm *via* atomic layer deposition (ALD) of 5 nm or 7.5 nm $SiO_2$ respectively [39]. For an intermediate size, the templates with 35 nm pore diameter were used without additional coating. A 30 nm gold layer was sputtered as a seed layer for electrodeposition on one side, followed by a gold deposition to increase the thickness. The individual, uncoupled Ni nanomagnets were grown from a Watts-type electrolyte containing 0.76 M nickel sulfate hexahydrate, 0.84 M nickel chloride hexahydrate and 0.49 M boric acid. The deposition is conducted potentiostatically in a three electrode cell with a deposition potential of -1.1 V *vs.* an Ag/AgCl/KCl reference electrode. An overview of the synthesized structures is shown in Figure 2.

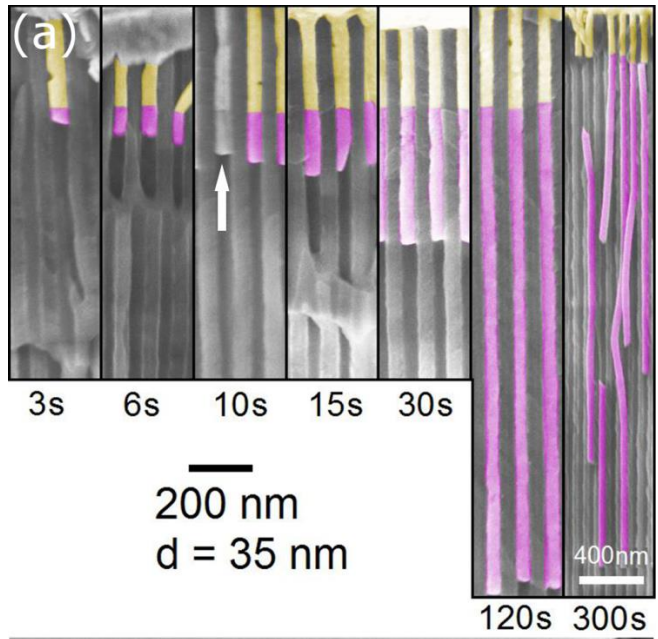

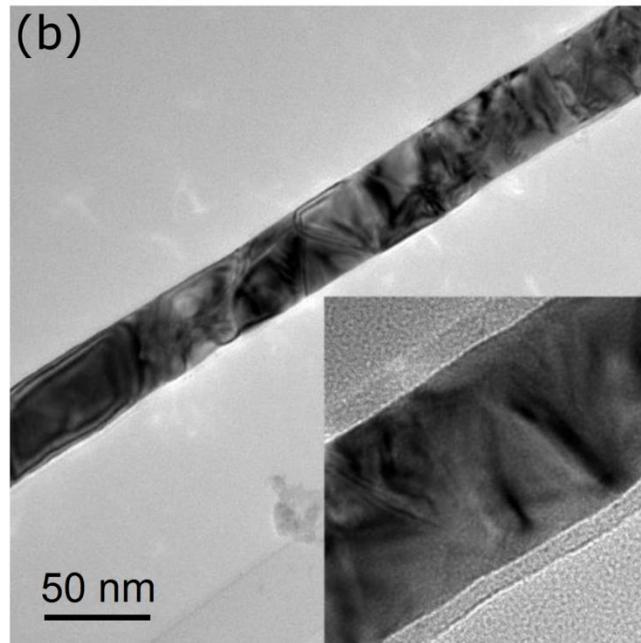

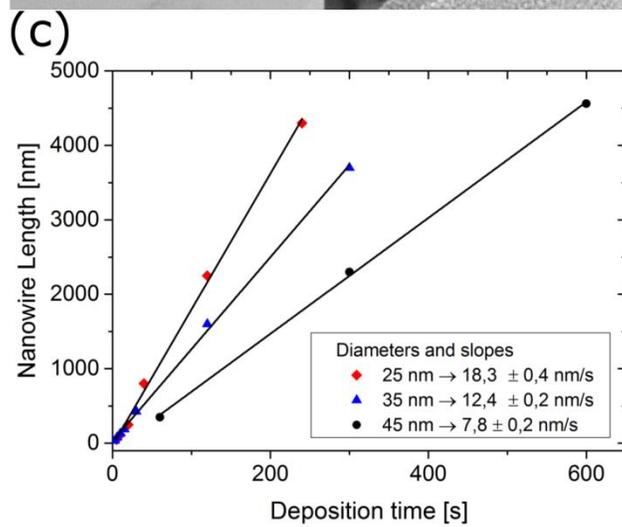

**Figure 2** (a) False color SEM images of electrodeposited Ni wires (purple) with the Au seed layer (gold) seen in the cross section of an AAO template. The arrow highlights an uncolored wire, demonstrating that Ni and Au can be distinguished from the SEM specific material contrast using an in-lens detector. (b) TEM image of a d = 45 nm nanowire displaying a polycrystalline structure. The inset shows a magnification and the bright outer $SiO_2$ layer. (c) The average wire length is measured over more than 50 wires as a function of deposition time. As expected, the relation is linear, indicating no diffusion constraints [40].

The temperature is left at room temperature and an additional stirring was not found to be necessary. In preliminary thin-film optimization experiments, the deposits were found to be homogenous and mirroring. We do not observe any difference in the electrodeposition conditions between $SiO_2$-ALD-coated and uncoated membranes.

*2.2 Synthesis of segmented nanowires.* In order to couple the nanomagnets, multi-segmented wires consisting of Ni and Cu segments with varying length were grown, as schematically shown in Figure 3a. It has been shown that segmented Ni/Cu nanowires can be grown with well-defined interfaces and morphologies [28,29]. In order to allow for a voltage specific deposition of high quality materials, a different electrolyte is chosen, containing 0.02 M copper sulfate pentahydrate, 0.1 M sodium sulfate, 0.25 M boric acid, 0.25 M sulfamic acid and 0.74 M nickel sulfate hexahydrate at a pH of 3.25. The copper deposition potential is at -0.5 V *vs.* an Ag/AgCl/KCl reference electrode, whereas nickel is deposited at -1.1 V. Backgrounds of the optimization of the electrolyte can be found in publications of Toth *et al.* [41,42]. The wires used in this study are designed to be released from the host template and characterized individually *via* electric transport measurements. During the device fabrication process it was found that multi-segmented wires with µm-long copper segments could not be released from AAO templates intact, but instead large voids and oxidized parts would hamper any measurements of electric transport. Several acids and bases of varying strength (chromic acid, phosphoric acid and sodium and potassium hydroxide) were investigated, however with identical, negative results. We consider this an interesting observation, because in former experiments the dissolution of Ni/Co/Cu wires using identical electrolytes has been successful [27]. The difference, however, is the thickness of the Cu segments, which is below 10 nm in the successful cases. It appears the close proximity to the less noble magnetic materials, which for themselves produce a passivating oxide layer, stabilizes the Cu segments. We solve this issue by using porous polycarbonate membranes synthesized using ion-track etching [43], which can be dissolved in organic solvents (Di/Trichlormethane) that have no effect on the metallic nanowire parts. The diameters of these membrane pores are 50 nm which is reduced by 5 nm $SiO_2$ deposition by ALD to 40 nm. The electrochemical conditions stay the same. Note that in this case the copper parts tend to oxidize over the course of a few days at ambient air as well, but sample storage under $10^{-4}$ mbar vacuum in dry atmosphere was found to be sufficient to slow down the oxidation. Energy dispersive X-ray spectroscopy (EDS) on newly grown samples within the membrane reveals a material composition of approximately 90% Ni and 10% Cu in the magnetic parts and not further quantifiable traces <3% of Ni in the Cu parts.

*2.3 Synthesis of single nanowire devices for measurements of the electric transport properties.* The nanowire templates are dissolved as discussed above and filtered using a NdFeB magnet to hold the wires in place in a thin vial while the liquid is exchanged with water and then with ethanol several times. A few drops of the ethanol suspension is spread on a Si/SiO$_2$ (300 nm) wafer, dried at ambient air and then rinsed again with ethanol. The targeted density of nanowires is around one nanowire per 1000 µm². The wafer is spincoated with a lift-off resist (LOR-3B MicroChem) and a positive photoresist (ma-P 1205 MicroChem). The exposure is conducted via direct lithography using a laserwriter (µPG 101 Heidelberg Instruments GmbH). Before the deposition of the Ti(2 nm)/Pt(50 nm) contact material, the SiO$_2$ oxide layer is removed in-situ using physical etching through an argon plasma. An SEM image of a contacted wire is shown in Figure 3b and a HR-TEM image in Figure 3c.

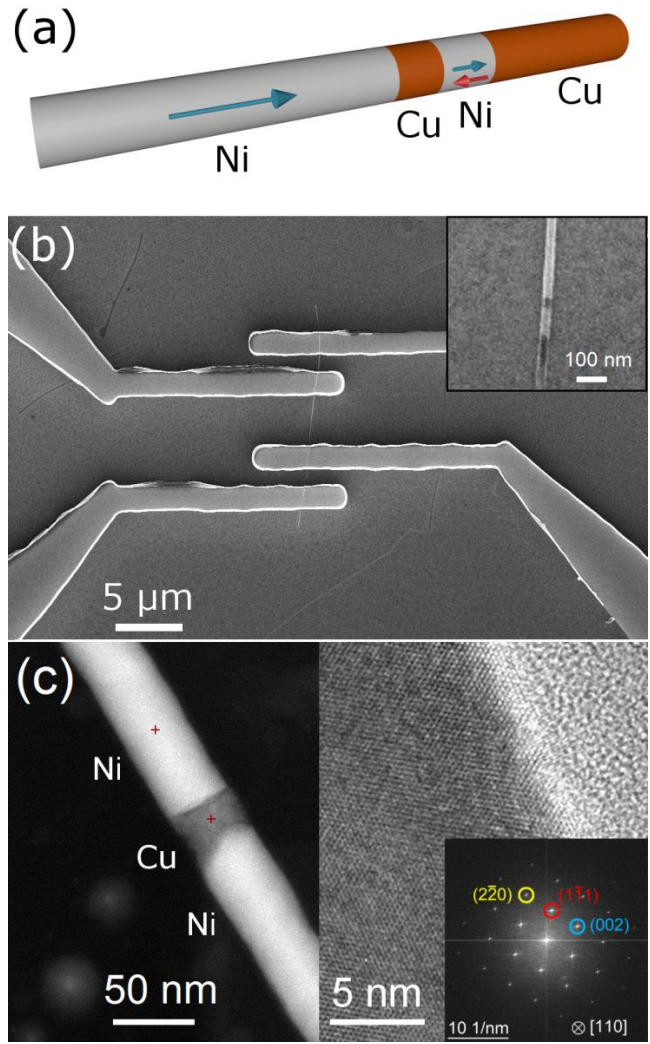

**Figure 3** (a) Schematic of coupled nanomagnets with a long and a short segment. (b) 4-point measurement device of a multi-segmented nanowire. The inset shows a magnification of the center region. The Cu segments (dark) have been selectively oxidized in weak KOH in order to enhance SEM contrast. (c) STEM image of a Ni(long)/Cu/Ni(short) interface region with the copper interlayer in dark contrast. The wire growth direction is from bottom right to top left. The red crosses denote the positions of EDS-elemental mapping shown in the supporting information. The right part shows a HRTEM of the Ni-parts, displaying a large crystalline region and a thin oxide layer (bright) on the outside. The fast Fourier transform showing the orientation of the crystalline Ni section is displayed as an inset.

## 3. Results and Discussion

*3.1 Properties of individual nanomagnets.* As discussed in the introduction, knowledge of the nanomagnet's individual magnetic properties is crucial for their later use. In a first step in order to investigate the coercive field dependence as a function of the nanomagnet length for each diameter, several different lengths were grown by adjusting the deposition time. The filled AAO-membranes were measured using vibrating sample magnetometry (VSM) (Versalab, Quantum Design Inc.) with $B$ parallel to the wire axis. Because in a VSM a large amount of wires is measured simultaneously, we acquire an average value for the coercivity. Previous investigations have shown that the coercive field of such an averaged hysteresis loop is only weakly dependent on the dipolar interaction field, arising from the close proximity of the nanowires to each other [14]. We intentionally chose not to synthesize multi-segmented nanowires to rule out a longitudinal coupling between the segments [29]. The raw hysteresis loops are shown in Figure SI-1 in the supporting information and Figure 4 summarizes the extracted data for three different diameters. There is a strong dependence of the coercive field on the nanomagnet length for short wires; however it is quickly saturating at about 1000 nm, as expected for nanoparticles that are dominated by shape anisotropy. Shape anisotropy can be described by the difference in the demagnetization factor $\Delta N$ between two magnetic axes with the shape anisotropy constant $k_{\text{shape}} = \frac{1}{2\mu_o} J_s^2 \Delta N$ and the anisotropy field $\mu_0 H_A = J_s \Delta N$, where $J_s$ is the spontaneous polarization [40]. We can neglect crystalline anisotropy because we have a polycrystalline material (see Figure 2b). A qualitative investigation of the dipolar interaction field using first order reversal curves (FORC) [44,45] for the 45 nm diameter wires, which are expected to have the highest interaction field, can be found in the supporting information.

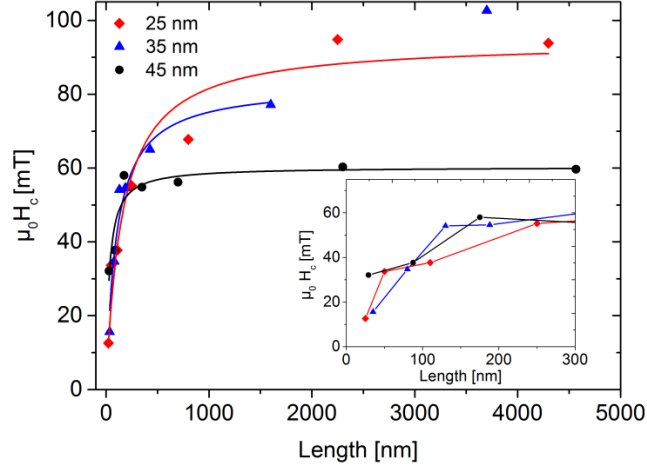

**Figure 4** Coercive field vs nanowire length for three 25 nm, 35 nm and 45 nm diameter nanowires. The lines are hyperbolic fits. The longest d = 35 nm nanowire sample shows an unexpectedly large coercivity value at 3700 nm, that is not incorporated into the fitting. Inset: Magnification of L < 300 nm.

Our results show that a significant change in the coercive field can be induced simply by changing the nanomagnet lengths. The smallest segments with an aspect ratio around 1 may, however, not exhibit sufficient anisotropy to be bistable along the wire axis. In order to resolve the involved reversal modes in the switching process, we conduct Monte Carlo simulations. The diameter of the simulated segments is 25, 35 and 45 nm with lengths of 29, 88, 175, 350 and 500 nm. The internal energy $E_{tot}$ of a nanowire with R magnetic moments can be written as:

$$E_{tot} = -\sum_{i<j}^{R} J\vec{m}_i \cdot \vec{m}_j + \sum_{i<j}^{R} \frac{\vec{m}_i \cdot \vec{m}_j - 3(\vec{m}_i \cdot \hat{n}_{ij})(\vec{m}_j \cdot \hat{n}_{ij})}{r_{ij}^3} - \sum_{i}^{R} \vec{H} \cdot \vec{m}_i$$

The first term is the exchange energy, the second term is the dipolar energy and the last term is the Zeeman energy. In this expression J=160 T/µB is the exchange coupling constant between nearest neighbors and $\vec{m}_i$ is the magnetic moment vector with a value of 0.615 µB. µB is Bohr's magneton. The Ni atoms are distributed in a fcc structure with a lattice parameter of 0.352 nm. Then, $r_{ij}$ is the distance between two magnetic moments $\vec{m}_i$ and $\vec{m}_j$, and $\hat{n}_{ij}$ is the unitary vector that connects the position of the atom i with the atom j. Finally, $\vec{H}$ is the applied magnetic field.

The Monte Carlo simulations were done by using the Metropolis algorithm with local dynamics and single-spin flip methods [46]. The new magnetic moment orientation is chosen randomly with a probability $p = \min[1, \exp(-\Delta E/k_B T)]$, where ΔE is the energy difference due to the variation of the magnetic moment, $k_B$ is the Boltzmann constant and T = 300 K. We did hysteresis loops with a saturation field of 200 mT. The applied field was varied in steps of *ΔH* = 1 mT. In every step of the field, the related value of the normalized magnetization of the system was determined after 3000 Monte Carlo steps. The results were obtained by doing at least five independent realizations. In order to reduce the calculation time due to the large number of magnetic moments in each wire, we used a scaling technique combined with Monte Carlo simulations, explained in detail in the references [47,48]. Following this technique we used χ=0.002 and $\eta \approx 0.55$, the exchange constant was scaled to $J' = J\chi = 0.32 \text{ T}/\mu\text{B}$, the temperature was scaled to $T' = T\chi^{3\eta} = 0.011 \text{ K}$ and the geometrical parameter were scaled as $L' = L\chi^{\eta}$ and $d' = d\chi^{\eta}$. With these scaled parameters the size of the scaled system was reduced. Consequently, the number of magnetic moments in each wire was decreased so that $R < 3000$, making the Monte Carlo calculations feasible in reasonable computational times.

The micromagnetic configurations of the d = 25 nm and l = 25 nm nanomagnet extracted from the Monte Carlo simulations is shown for different field intervals in Figure 5a. The shortest d = 35 nm nanomagnets show identical results. In the d = 25 and d = 35 nm case, we find magnetization reversal through coherent rotation. Interestingly, our results indicate that the d = 45 nm and L=25 discs are large enough to host a vortex state (Figure 5b), which explains why the coercive field of the shortest d = 45 nm nanomagnets are higher than for the other diameters. Hence, for the lowest aspect ratios, a bistable configuration cannot be found.

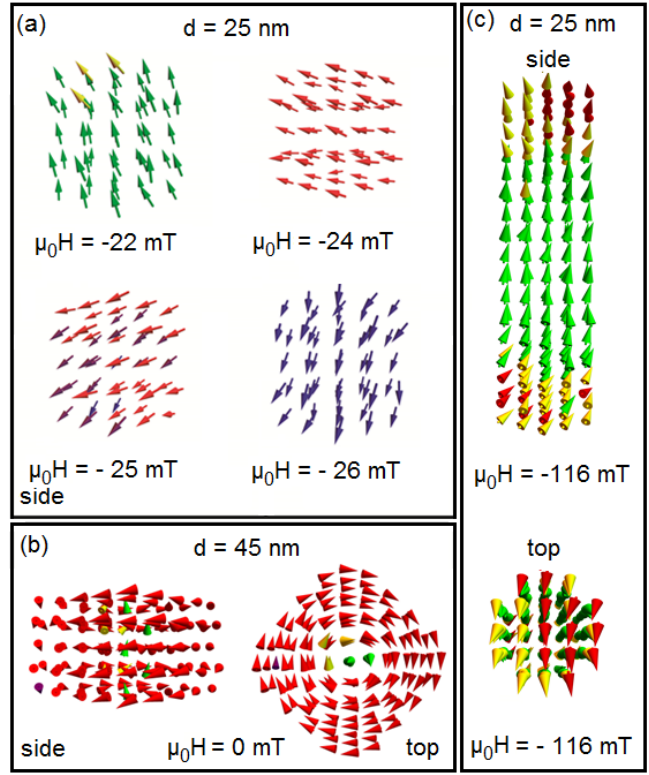

**Figure 5** Selected Monte Carlo simulations of nanomagnets with varying length and diameter. (a) d = 25 nm nanomagnet with an aspect ratio near 1 displaying coherent rotation. (b) d = 45 nm nanomagnet with an aspect ratio near 1 displaying a vortex state at zero field. (c) d = 25 nm nanomagnet with an aspect ratio of 4 seen from the side and from the top. The nucleation process of a transverse domain wall can be seen. Increasing the magnetic field strength by -2 mT will lead to the propagation of the domain wall, i.e. a rapid switching process, as expected [49,50].

At an aspect ratio of 2, the Monte Carlo simulations indicate that the segments are bistable. Therefore, this ratio will be defined as the nanomagnets lower length limit. Figure 5c shows the magnetization reversal through a transverse domain wall. Note that the long d = 45 nm nanowires lie at the transition region between vortex and transverse domain walls. It is known that Ni wires may shift their reversal mode from vortex domain walls to transverse domain walls at a critical diameter of approximately 42 nm [51,52].

The strong dependence of the coercive field for short lengths is due to a high change of the z-component $N_z$ of the demagnetization tensor $\overleftrightarrow{N}$, which is defined through the demagnetizing field $\vec{H_D} = \overleftrightarrow{N}\,\vec{M}$ and the magnetization vector $\vec{M}$. All calculated $N_z$ values are listed in Table 1. Longer segments hardly show any significant change, because the reversal mode is *via* the propagation of a domain wall that conserves the magnetic arrangement (most of the moments parallel to the long wire axis) during the process.

Table I: Calculated components of the demagnetization tensor $\overleftrightarrow{N}$ for the simulated lengths

| Length [nm] | d = 25 nm | | d = 35 nm | | d = 45 nm | |
|---|---|---|---|---|---|---|
| | $N_z$ | $N_x = N_y$ | $N_z$ | $N_x = N_y$ | $N_z$ | $N_x = N_y$ |
| 25 | 0.302738 | 0.348631 | 0.311577 | 0.344212 | 0.413085 | 0.293458 |
| 50 | 0.175693 | 0.412154 | 0.162296 | 0.418852 | 0.185333 | 0.407334 |
| 110 | 0.086684 | 0.456658 | 0.105285 | 0.447358 | 0.100936 | 0.449532 |
| 250 | 0.039592 | 0.480204 | 0.074699 | 0.462650 | 0.052505 | 0.473747 |
| 800 | 0.012619 | 0.493690 | 0.033948 | 0.483026 | 0.026765 | 0.486616 |
| 2250 | 0.004512 | 0.497744 | 0.009224 | 0.495388 | 0.008256 | 0.495872 |
| 4300 | 0.002365 | 0.498818 | 0.004003 | 0.497999 | 0.001467 | 0.497915 |

In order to further elucidate the reversal mechanisms, we conduct angle-dependent measurements of the coercive field using a VSM. The data is acquired by taking a full hysteresis loop with the magnetic field applied in 10° steps as can be seen in Figure 6. The shortest segments (red line) clearly show no significant angle dependency, indicating that no stable magnetic configuration can be found. At an aspect ratio of approximately 2 (orange line), the coercive field is at its highest value for parallel magnetic fields (0°), because an easy axis establishes. The angle dependent coercive field trend of long magnetic nanowires has been thoroughly studied before [53-55]. Their theoretical model is well represented in our data by the smooth decrease of the coercive field with increasing angle. We do not observe any qualitative difference between all diameters that could hint towards a transition from a transverse to a vortex domain wall, as indicated by the simulations.

We additionally conducted single-wire measurements of the magnetoresistance for each wire as presented in Figure 6d and e. From the AMR curves the switching field is extracted and plotted against the angle. A slight asymmetry is present which is due to a possible misalignment. Additionally, we observe an anomaly with a local maximum near $B||I$ and a minimum around 30 degrees. Such a behavior has been observed before by Wernsdorfer *et al.* [56], suggesting it could be understood as a reminiscence of a Stoner-Wohlfarth-like [57] switching behavior. However, all AMR curves (Figure 6d) all display a distinct, sharp jump upwards, which is a clear indication of a nucleation and propagation process [58]. We additionally synthesize and measure d = 70 nm nanowires under identical conditions, which does not show the anomaly (Figure 6d, green line). It might therefore be a peculiarity which is due to the magnetization reversal via a transverse domain wall, because a slightly tilted applied field might facilitate the domain wall's nucleation.

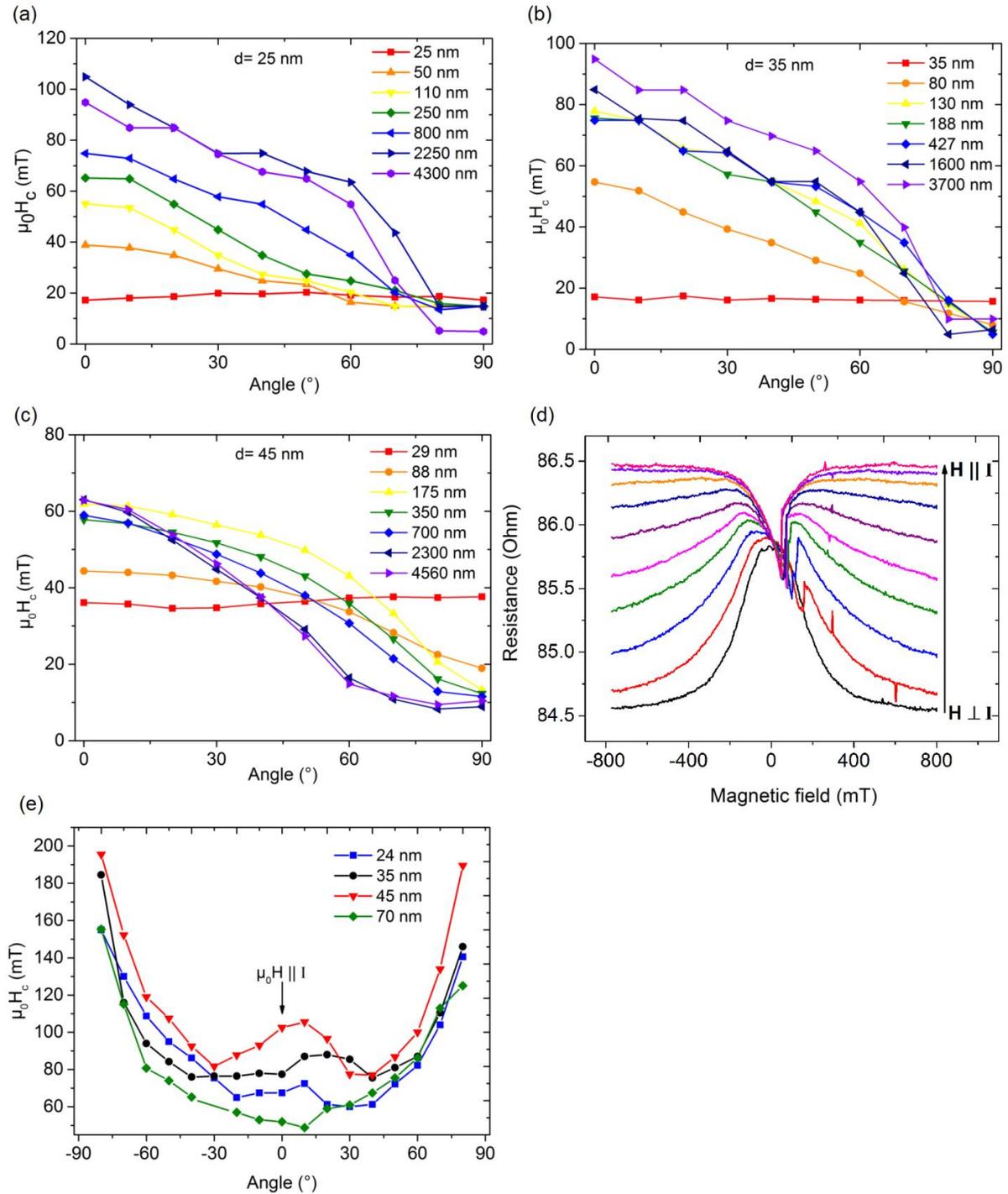

**Figure 6** (a)-(c) VSM measurements of the coercive field as a function of the applied field angle. The shortest nanomagnets show no real angle dependency, indicating insufficient anisotropy for a bistable magnetic configuration. (d) shows exemplary AMR curves (45 nm) of the longest nanowires (4560 nm) and (e) shows the angle dependency of the switching field as a function of the applied outer field angle for the longest nanowires of each diameter.

Note that in contrast to the coercive fields determined by VSM (Figure 4), the single 45 nm nanowire has the highest overall coercive field. Drawing any conclusions about the absolute coercive field strength would be elusive, because only one nanowire of each kind was measured in this case and the switching field distribution of single wires spread on a wafer can be surprisingly large due to defects and irregularities [14]. The nanowires show linear, metallic R vs T behaviors with residual resistivity ratios of $\rho(300K)/\rho(2K) \approx 4.5$, which is a notably high value for electrodeposited, ferromagnetic nanowires [27,36,59,60], indicating a good sample quality (see supporting information).

*3.2 Nanowires with coupled segments.* Based on the discussion and evaluation in Section 3.1, we chose nanomagnet lengths of 5 µm for the long Ni segment, 20-100 nm for the Cu length and an aspect ratio of 2.5 for the short Ni segment, in order to allow for a sufficient difference in coercive fields (Figure 4). The diameter of the coupled nanowires is 40 nm. The length of the first segment is comparably large, because it is used for the contacting procedure with a 4-point contact. Measurements of the magnetoresistance (MR) are conducted using a *PPMS Dynacool* (Quantum Design Inc.) between 2.5 K and 350 K. Both segments are saturated with a magnetic field that is applied parallel to the nanowire axis. The resistance is measured while the magnetic field is swept with a rate of 1 mT/s. Figure 7a displays the raw data without the subtraction of any kind of background of typical MR-curves obtained for $L_{Cu}$ = 100 nm at 170 K and 2 K. We chose to display 170 K, because the GMR signal originating from only two segments is notably small and therefore in all the cases we see a superposition of signal originating from anisotropic magnetoresistance (AMR), as it is expected from a long nanowire, and a giant magnetoresistance signal, which occurs as soon as both segments are aligned antiparallel. The curve at 2 K alone would not evidence the existence of the antiparallel state, because it could be thinkable that the long segment reverses partly. Therefore a first step would become visible in the MR curve, which would be purely due to an AMR effect. The fact that the curve undoubtedly jumps above the zero line for at least one temperature clearly evidences the existence of a GMR effect which can only be due to an antiparallel alignment of both segments. The absolute signal strength is notably small, which can be caused by the fact that the copper segments are not entirely pure due to a nickel contamination, which decreases the spin diffusion length, which can be as large as 1 µm in Cu [61]. The temperature dependence of the GMR signal behaves as expected. Figure 7b and shows increasing absolute signal strength with decreasing temperature, caused by a freeze-out of scattering sites. Additionally, decreasing the copper layer length to 30 nm increases the signal strength as expected, further reinforcing our claim that the observed jumps are due to the GMR effect. Interestingly, even further reduction of the copper segment length does not lead to an increase of the GMR signal strength, but to its vanishing at temperatures below 280 K. In this case we only observe one single jump, in contrast to all other cases, where always two jumps are present. This effect can be explained by the coupling strength: If the coupling between the nanomagnets is too strong, they behave as one and switch simultaneously. If the temperature is increased, the saturation magnetization is slightly lowered (note that Ni has a low Curie temperature of only 636 K), which leads to lower stray fields. Above 280 K a GMR signal reappears, that is very narrow on the magnetic field axis $H$. During a field sweep from negative to positive saturation, the higher the stray field from the long segment, the higher is the effective magnetic field that is felt by the short segment, which means that it tends to switch late. We can therefore compare the width $\Delta H$ of the region in which the GMR-signal is present as a function of the copper interlayer thickness. At 300 K we find widths $\Delta H$ of 54 mT, 15 mT and 7 mT for 100 nm, 30 nm and 20 nm respectively, which supports our thesis of stronger coupling with narrower distance. Misalignments, bent nanowires or defects may lead to an L-state between both segment's magnetizations, which explains why the absolute strength for 20 nm at T > 300 K does not follow the trend.

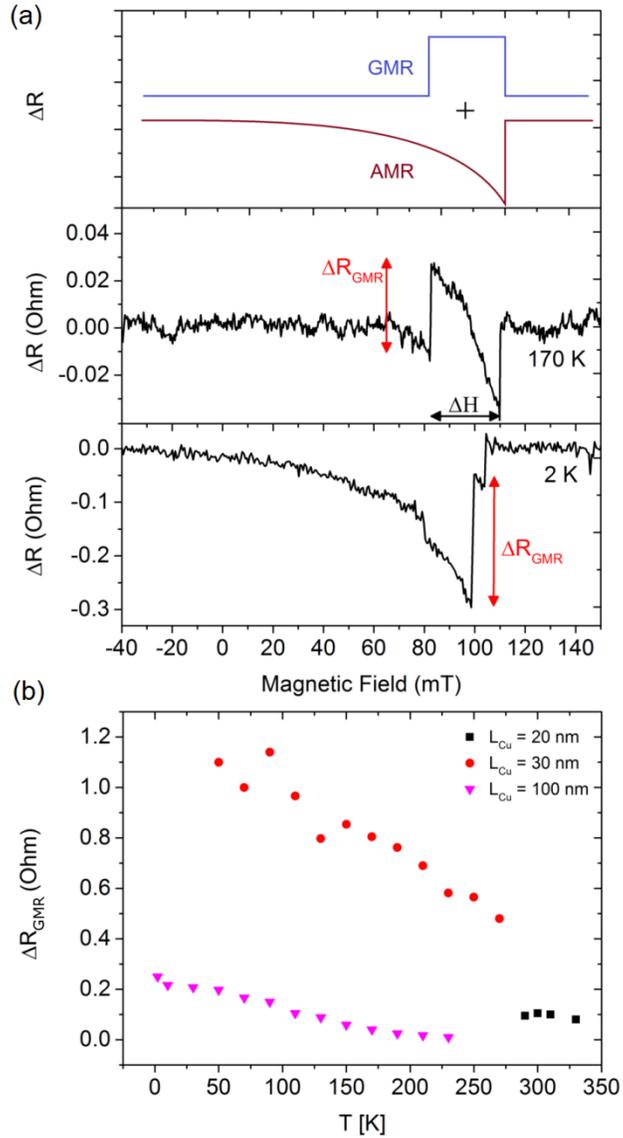

**Figure 7** GMR signal measured in nanowires (d = 40 nm) consisting of coupled nanomagnets. The length of the long and short magnetic segments is 5 μm and 100 nm, respectively. (a) Raw data without any background subtraction from the same wire at 170 K and 2 K shows a typical field dependence of the resistivity: a superposition of an AMR and a GMR signal. The AMR signal (red) originates from the long Ni segment and the GMR signal (blue) originates from an antiparallel alignment of the long and the short Ni segment. Both signals are additive, because they occur in different wire segments, i.e. in series. Note that both AMR and GMR signals have different temperature dependence, which is why the first GMR step does not always jump over the base line. ΔH is the difference in coercive field of both segments, however in a coupled configuration. The behavior is perfectly reversible and symmetric in both sweeping directions (not shown). A magnetic force microscopy (MFM) image is of a nanowire in parallel configuration is shown in the supporting information (SI-6). (b) The height of the first jump, i.e. the resistance contribution due to the GMR effect, is plotted as a function of temperature.

*3.3 Analytical model.* Based on the analysis in section 3.2, the question arises what is the minimum distance between two magnetic segments, if an antiparallel state shall be achievable. Naturally, this distance will depend on the geometric and magnetic properties of both involved segments. We therefore develop a simple, analytical model for the determination of the stable antiparallel switching regimes. The investigated system and its parameters are introduced in Figure 8.

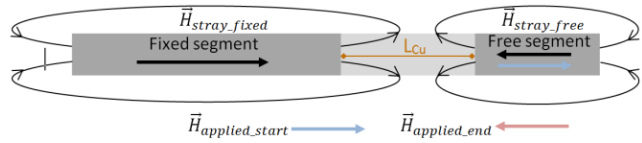

**Figure 8** Geometry of the modelled system in the antiparallel state. Each magnetic segment has its own coercive field and stray field.

The stray field of an arbitrary cylinder along its center axis $z$ can be calculated from solving the Poisson equation. It is given by

$$\mu_0 H(L_\mathrm{m}, z) = \frac{B_\mathrm{r}}{2}\left[\frac{L_\mathrm{m} + z}{\sqrt{r^2 + (L_\mathrm{m} + z)^2}} - \frac{z}{\sqrt{r^2 + z^2}}\right] \qquad (1)$$

with the remanence induction $B_\mathrm{r} = 640$ mT for nickel, the cylinder radius $r$ and the magnetic segment length $L_\mathrm{m}$. $z = 0$ corresponds to the surface of the cylinder edge. Both segments create a stray field which interacts with the corresponding other segment. For example, the stray field created by the fixed segment can nucleate a switching process of the free segment, if it is large enough. In order to nucleate a switching process in the free segment that is caused by the stray field from the fixed segment, a sizeable amount (nucleation volume) of the free segment needs to be switched by the interaction field. The question arises, at which position the value of the stray field should be considered. It is obviously not the center of the free segment in a point-dipole approach, as magnetization reversal is not coherent for the dimensions considered. However, the reversal occurs through the growth of a nucleation volume at the wire end. For a nanomagnet aspect ratio of 2.5 or larger, the extent of this volume reaches a plateau, which is comparable to the fraction of the wall width [62]. In the case of transverse domain walls (between 3 and 10 times the exchange length in nickel), the domain wall width is known to be twice the wire diameter [63-65]. We therefore introduce a coordinate transform with an effective distance $z'$ calculated from the sum of gap size $L_{cu}$ plus an additional length reflecting the nucleation volume: $z \to z' = L_{cu} + \alpha \cdot d$, where $\alpha$ is a phenomenological parameter on the order of 1 and $d$ the diameter. In order to check the validity of this picture and to provide a numerical value for $\alpha$, we performed a series of micromagnetic simulations using the feeLLGood simulation package [66]. Two cylinders of infinite length (mimicked by boundary conditions canceling end charges), various diameters and gap were considered, and the switching field in the case of initial head-to-head configuration was calculated. Eq.(1) applied to z' can be reworked to in the form of a scaling law:

$$\sqrt{\frac{3}{4}\left(\frac{M_s}{3H_{stray}} - 1\right)} = \frac{L_{cu}}{r} + 2\alpha \qquad (2)$$

The simulated variation of $H_{stray}$ versus $(L_{cu}/r)$ is found to have a slope very close to 1, which hints at the reliability of the model. The corresponding diagram is shown in the supporting information in Figure SI-5. Besides, the value of $\alpha$ can be inferred as the intercept with the y-axis. In simulations this intercept only weakly depends on r, with $\alpha \simeq 0.5$. Hence the effective gap size $z'$ corresponds to the copper layer thickness plus half the wire's diameter.

If two nanomagnets are brought into close proximity to another, their interplay is not only defined by their distance and stray field as discussed above, but also by the size of their coercive field and an applied external field. In order to take all of these parameters into account, we develop phase diagrams (Figure 9). In these phase diagrams we set the fixed segment length to 1 µm, because all relevant parameters such as stray field strength and coercive field saturate from there on. The corresponding coercive field of the fixed segment is derived from Figure 4. The coercive field of the free segment is a free parameter plotted on the y-axis; however it is linked to the segment's length, again *via* the relation described in Figure 4. Note that the length dependence of the stray field strength is also accounted for through Eq.(1).

We consider a starting case in which both segments are saturated and aligned in the same direction. The magnetic field is then swept towards increasingly negative values. As already discussed, an antiparallel state is magnetostaticly unfavorable. For its creation, three different criteria need to be fulfilled:

1. If the stray field of the fixed segment is larger at a certain distance $z$ than the nucleation field of the free segment, they will always couple and behave as one segment. Thus, we need the condition:

$$H_{c\_free} \overset{!}{>} H_{stray\_fixed}$$

2. In a state where both segments are parallel, the fixed segment stray field stabilizes the free segment. Therefore the external applied field during the field sweep needs to be larger than the sum of the coercive field of the easy segment plus the stray field of the fixed segment, or else the segment will not switch.

$$H_{\text{applied}} \overset{!}{>} H_{\text{c\_free}} + H_{\text{stray\_fixed}}$$

3. The external field may not be so large that the fixed segment switches after the easy segment has switched, which creates its own stray field.

$$H_{\text{c\_fixed}} \overset{!}{>} H_{\text{applied}} + H_{\text{stray\_free}}$$

For a given geometry, if a magnetic field exists, in which all three criteria are fulfilled, the cells are colored blue and an antiparallel configuration is theoretically possible. Figure 9 displays the results for three different diameters. The coercive field of the free layer shown on the vertical axis is determined as a function of the segment length from the hyperbolic fits in Figure 4.

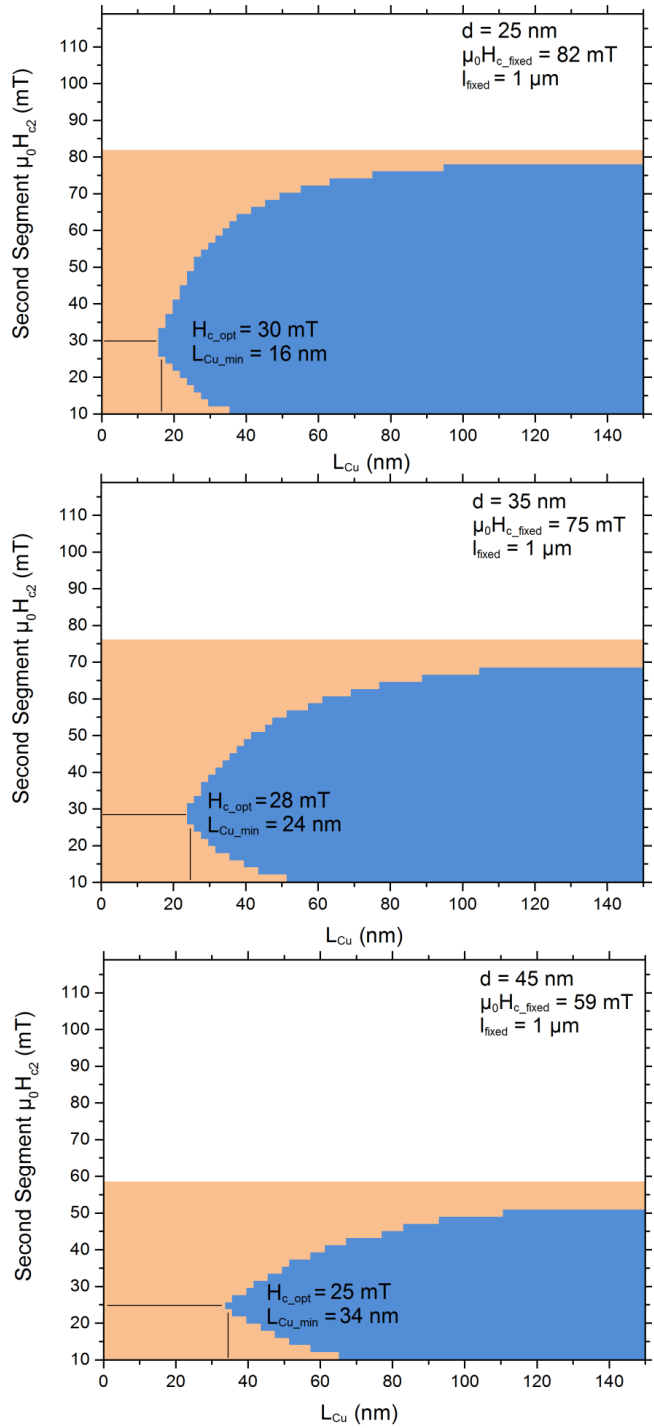

**Figure 9** Phase diagrams for d = 25 nm, d = 35 nm and d = 45 nm nanowires with the minimum gap size between two segments of 16 nm, 24 nm and 34 nm, respectively. The blue (red) regions denote configurations of minimum distance and free layer coercive field where an antiparallel configuration is possible (not possible). Note that the 45 nm case is modelled using a transverse wall, even though micromagnetic simulations indicate a vortex wall may be present.

From the phase diagrams, ideal coercive field values $\mu_0 H_{c\_opt}$ for the free layers can be inferred, because at that value the segment distance can be minimal. These are 30 mT, 28 mT and 25 mT for diameters of 25 nm, 35 nm and 45 nm, respectively.

The wires that were used in the GMR experiments have a diameter of 40 nm. For this case, interpolation between the phase diagrams for 45 nm and 35 nm indicates that the minimum distance of two segments is below 30 nm. This finding reinforces the interpretation of the data presented in Figure 7 in which a gap of 20 nm (at d = 40 nm) is too narrow to host a real antiparallel state obtained through sweeping the external field.

**Conclusion**:

The above assessment of electrodeposited longitudinal soliton geometries shows that the individual properties of electrodeposited nanomagnets can be precisely tuned. We investigate the magnetization reversal as a function of the nanomagnets diameter and aspect ratio and identify a transition from coherent rotation to transverse domain walls. We show that nanomagnets can be coupled with different strengths as a function of the distance between each other. It is proved that individual magnetic segments can be switched through external magnetic fields, giving rise to a GMR effect. We develop an analytical model that defines regimes, in which an antiparallel state between two nanomagnets is possible. The most important step towards the development of working longitudinal soliton shift registers will be the development of a data propagation mechanism. In future studies we will attempt to increase the GMR signal for spin-valve applications and conduct a thorough investigation of the switching behavior using MFM.

**Acknowledgements**

The research leading to these results has received funding from the European Unions's 7th Framework Programme under grant agreement n°309589 (M3d). We further acknowledge support from FONDECYT 1160198 and 1161018, Financiamiento Basal FB 0807 para Centros Científicos y Tecnológicos de Excelencia.